\documentclass[11pt]{article}
\usepackage[T1]{fontenc}
\usepackage{amsfonts}
\usepackage{amsmath}
\usepackage{amssymb}
\usepackage{amsthm}
\usepackage{graphicx}
\usepackage{braket}
\usepackage{typearea}
\usepackage{fullpage}

\newcommand{\CC}{\mathbb{C}}
\newcommand{\QQ}{\mathbb{Q}}
\newcommand{\RR}{\mathbb{R}}
\newcommand{\comp}{\text{comp}}
\newcommand{\defeq}{\mathrel{:=}}

\newcommand{\isom}{\cong}
\newcommand{\matrixring}{M}
\newcommand{\op}{\operatorname{op}}

\newcommand{\Sym}{\operatorname{Sym}}
\newcommand{\tensor}{\otimes}

\def\Ltot{L^\mathrm{tot}}

\def\mytilde{\widetilde}
\def\LU{LU}
\def\NP{\text{\textsc{NP}}}
\def\borderrank{\mathop{\underbar{\ensuremath R}}}
\newcommand{\veps}{\varepsilon}
\newcommand{\vphi}{\varphi}

\def\eqbd{\mathop{{:}{=}}}

\def\C{\mathbb{C}}

\def\F{\mathbb{F}}
\def\H{\mathbb{H}}

\def\om{{\rm emax}}

\def\pre{\hbox{\small\sc Pre}}
\def\post{\hbox{\small\sc Post}}

\newcommand{\ve}{\varepsilon}

\theoremstyle{definition}
\newtheorem*{definition}{Definition}
\newtheorem*{example}{Example}
\newtheorem*{notation}{Notation}
\newtheorem*{theorem}{Theorem}
\newtheorem*{remark}{Remark}
\newtheorem*{corollary}{Corollary}

\def\Cohnetal{Cohn~\textit{et al.}}
\begin{document}

\title{Computational Complexity and Numerical Stability \\ of Linear Problems}
\author{Olga Holtz
\and Noam Shomron
}
\date{\small Jun 18, 2009}

\maketitle

\begin{abstract}  \noindent We survey classical and recent developments in numerical linear algebra,
focusing on two issues: computational complexity, or arithmetic costs, and numerical stability,
or performance under roundoff error.  We present a brief account of the algebraic complexity
theory  as well as the general error analysis for matrix multiplication and related problems.
We emphasize the central role played by the matrix multiplication problem and discuss 
historical and modern approaches to its solution.
\end{abstract}

\bibliographystyle{alpha}


\section{Computational complexity of linear problems}
\label{sec:bilinear_complexity}

In algebraic complexity theory one is often interested
in the number of arithmetic operations required to
perform a given computation,
modelled as a programme which receives an input
(a finite set of elements of some algebra)
and performs a sequence of algebra operations
(addition, subtraction, multiplication, division,
and scalar multiplication).
This is called the total (arithmetic) complexity
of the computation.\footnote{This model
only counts the operations and
completely ignores storage and communications costs,
limitations on precision,
details of how arithmetic is implemented (in particular we are
not counting \emph{bit} operations), etc.}
Moreover, it is often appropriate to count only
multiplications (and divisions), but not additions
or multiplications by fixed scalars.  These notions
can be formalized \cite[Definition~4.7]{BCS}.
For now, let us invoke
\begin{notation}
Let $\F$ be a field, let $\F[x_1,\ldots,x_n]\subseteq A \subseteq \F(x_1,\ldots,x_n)$ be an $\F$-algebra, and let
$\Phi=\{\vphi_1,\ldots,\vphi_m\}$ be a finite set of functions.
The total arithmetic
complexity of~$\Phi$ will be denoted~$\Ltot_A(\Phi)$,
and its multiplicative complexity by~$L_A(\Phi)$.
\end{notation}
\noindent
Intuitively, this is the minimal number of steps required to compute
all of
$\vphi_1$, \dots,~$\vphi_m$
starting from a generic input~$(x_1,\ldots,x_n)$, with intermediate results in~$A$
(in all cases we consider, $A$ will simply be the algebra
of polynomials or of rational functions in the input variables,
and will not always be explicitly indicated).
The input and all intermediate results are understood to be stored in memory, and the simultaneous
computation of a set~$\Phi$ of functions means that at the end of the
programme $\Phi$ is contained in the set of results.\footnote{For example, $X_0 \defeq x$, \ $X_1 \defeq X_0\cdot X_0$, \ $X_2 \defeq X_1\cdot X_1$, \ $X_3 \defeq X_2\cdot X_0$ is a programme, using $3$~operations---solely multiplications in this example, which computes $x^5$, as well as any subset of $\{x, x^2, x^4, x^5\}$, in~$A=\F[x]$.}

Let $U$, $V$, and~$W$ be finite-dimensional vector spaces over~$\F$,
and consider the class of bilinear functions~$\vphi\colon U\times V \to W$
(which includes matrix multiplication).
To define the multiplicative complexity of such a function, choose
bases $\{u_i\}_{1\le i \le m}$, $\{v_j\}_{1\le j \le n}$, and~$\{w_k\}_{1\le k\le p}$, so that
\[
 \vphi\left(\sum_{i=1}^m x_i u_i, \sum_{j=1}^n y_j v_j\right)
 = \sum_{k=1}^p \vphi_k w_k.
\]
We regard the coefficients as variables, so that each~$\vphi_k$
is a homogeneous polynomial of degree~$2$ in~$x_1,\ldots,x_m,y_1,\ldots,y_n$.
\begin{definition}[\protect{Cf.~\cite[Definition~14.2]{BCS}}]
  $L(\vphi) = L_{\F[x_1,\ldots,x_m,y_1,\ldots,y_n]}\bigl(\{\vphi_1,\ldots,\vphi_p\}\bigr)$.
\end{definition}
\noindent
Because we are considering the multiplicative complexity, this is a well-defined notion that does not depend
on the choice of bases.

It turns out that the multiplicative complexity of a bilinear
function~$\vphi\colon  U\times V \to W$ is controlled by
a somewhat more well-behaved notion, the \emph{rank}~$R(\vphi)$.
This is a standard notion in multilinear algebra, which
generalizes that of the rank of a linear map.
\begin{definition}
Let $t\in V_1\tensor\cdots\tensor V_n$.  The rank~$R(t)$ is
the smallest~$r$ such that one can write
$t=\sum_{i=1}^rt_i$ with each~$t_i$ a monomial tensor, i.e., of the form $t_i=v_1\tensor\cdots \tensor v_n$ for some $v_i\in V_i$.
\end{definition}
\noindent
In case $\vphi\colon  U\times V\to \F$ is the bilinear map
corresponding to a linear function~$\mytilde{\vphi}\colon  U\to V^*$,
the rank~$R(\vphi)$ is the rank of~$\mytilde{\vphi}$ in the usual
sense.  Well-known algorithms, such as Gaussian elimination,
as well as the fast algorithms described in this paper
(see Section~\ref{sec:other_linear_problems}),
can quickly compute the rank of a matrix, but determining the
rank of a tensor of order~$3$ already seems to be quite difficult.
Computing the rank of a given tensor is a combinatorial or
algebro-geometric problem~\cite{L}.

We now explain how the rank controls the complexity of a
bilinear function.
First, by a known result of Strassen (see~\cite[Proposition~14.4]{BCS}),
if $\vphi\colon V \to W$ is a \emph{quadratic} map between
finite-dimensional vector spaces, that is,
\[
\vphi\left(\sum_{i=1}^n x_i v_i\right) = \sum_{j=1}^p\vphi_j(x_1,\ldots,x_n)w_j
\]
for some bases~$\{v_i\}_{1\le i\le n}$ (resp.~$\{w_j\}_{1\le j\le p}$) of~$V$ (resp.~$W$) and
homogeneous polynomials~$\vphi_i\in \F[x_1,\ldots,x_n]$ of degree two,
then we need not search through some rather large class of programmes
to find one which computes $\vphi$ optimally, for in fact $L(\vphi)=L_{\F[x_1,\ldots,x_n]}\bigl(\{\vphi_1,\ldots,\vphi_p\}\bigr)$ equals
the smallest~$l\ge1$ such that
\begin{equation}
  \label{eq:quadratic_linear_complexity}
  \vphi(v) = \sum_{i=1}^l f_i(v)g_i(v)w_i
\end{equation}
for some linear functionals~$f_i,\,g_i\in V^*$.
(Note that such a formula immediately gives an obvious algorithm
computing~$\vphi(v)$ using only $l$~(non-scalar) multiplications.)

Now let $\vphi\colon U \times V \to W$ be a bilinear map
between finite-dimensional vector spaces.
This is covered by the preceding result of Strassen, since a bilinear
map~$U\times V \to W$ may be regarded as a quadratic map
via the isomorphism~$\F[U\times V] = \F[U]\tensor \F[V]$.
A \emph{bilinear algorithm} for~$\vphi$ amounts to writing
\begin{equation}
  \label{eq:bilinear_rank}
  \vphi(u,v) = \sum_{i=1}^r f_i(u)g_i(v)w_i
\end{equation}
for certain linear functionals $f_i\in U^*$, $g_i \in V^*$,
and~$w_i\in W$.
The minimum such~$r$ is the rank~$R(\vphi)$.
Note that the rank of~$\vphi$ is not necessarily the same as its
bilinear complexity, despite the superficially similar-looking formulae
\eqref{eq:quadratic_linear_complexity} and~\eqref{eq:bilinear_rank}.
However, by decomposing a linear functional~$f\colon U\times V\to \F$
as $f(u,v) = f(u,0)+f(0,v)$, one can see that
\[
  L(\vphi) \le R(\vphi) \le 2L(\vphi).
\]
It is often easier to work with the rank rather than the more subtle
notion of multiplicative (or total) complexity, and the above inequality shows
we do not lose much in doing so.

\subsection{Algebraic complexity of matrix multiplication}

The basic problem is to compute the (total or multiplicative)
complexity of multiplying two $n\times n$~matrices.
This is a difficult question whose answer is not at present known
for~$n=3$, for instance.

Matrix multiplication is a bilinear problem (see Section~\ref{sec:bilinear_complexity})
\begin{gather*}
  \vphi\colon \matrixring_{n\times n}(\F)\times \matrixring_{n\times n}(\F) \to \matrixring_{n\times n}(\F)\\
  (X,Y) \mapsto XY=\left(\sum_{l=1}^nX_{il}Y_{lj}\right)_{1\le i,j\le n}
\end{gather*}
whose corresponding tensor
will be denoted
\[
  \braket{n,n,n} \defeq \sum_{1\le i,j,k\le n}u_{ij} \tensor v_{jk} \tensor w_{ki}.
\]
For~$n=2$ Winograd
proved~\cite{W} that seven multiplications are required,
so $L\bigl(\braket{2,2,2}\bigr) = R\bigl(\braket{2,2,2}\bigr) = 7$,
but
for~$n=3$ even the rank is not known at present
(it is known that $19\le R\bigl(\braket{3,3,3}\bigr)\le 23$;
see \cite[Exercise~15.3]{BCS}, \cite{L}).

Instead of fixing~$n$, one considers the \emph{asymptotic}
complexity of matrix multiplication:
\begin{equation}
\label{eq:omega}
  \omega(\F) = \inf\Set{ \tau\in\RR |
    \Ltot_{\F[X_{ij},Y_{ij}]}\left(\Set{\sum_{l=1}^nX_{il}Y_{lj}|1\le
      i,j\le n}\right)= O(n^\tau)}
\end{equation}
so that $n\times n$~matrices with entries in~$\F$ may be multiplied using
$O(n^{\omega(\F)+\eta})$~operations,\footnote{Technically, division is
  not allowed, as the computation should be in~$\F[X_{ij},Y_{ij}]$, although
this is no restriction if $\F$ is an infinite field
(see~\cite[Remark~15.2]{BCS}).}
for every~$\eta>0$.

First of all, one can replace the total complexity in~\eqref{eq:omega}
by the multiplicative complexity or by the rank~\cite[Proposition~15.1]{BCS} and get the same exponent.
Second, $\omega(\F)$ is invariant under extension of scalars
\cite[Proposition~15.18]{BCS}, so it does not depend on the exact
choice of field~$\F$ (e.g., $\QQ$ versus $\RR$ or $\CC$), but
rather only on its characteristic, which is usually taken to be zero
(so $\omega$ denotes $\omega(\CC)$).

The value of~$\omega$ is an important quantity in numerical linear
algebra, as it determines the asymptotic complexity of not merely
matrix multiplication but also matrix inversion, various matrix
decompositions, evaluating determinants, etc. (see
Sections \ref{sec:other_linear_problems} and~\ref{sec:stable_other_linear_problems}).

An obvious bound is $2\le \omega \le 3$, since the straightforward
method of matrix multiplication uses $O(n^3)$~operations, on one hand,
while on the other hand we need at least $n^2$~multiplications to
compute $n^2$~independent matrix entries.
The first known algorithm proving that $\omega<3$ was
Strassen's algorithm, detailed in Section~\ref{sec:recursive_matrix_multiplication}, which starts with an algorithm for
multiplying $2\times 2$~matrices using seven multiplications and
applies it recursively, giving $\omega\le\log_27$.  This idea of
exploiting recursion will be explored in the next section.

\subsection{Asymptotic bilinear complexity via tensor ranks}

The basic idea behind designing fast algorithms to multiply
arbitrarily large matrices, thereby obtaining good upper bounds
on~$\omega$, is to exploit recursion: multiplication of large matrices
can be reduced to several smaller matrix multiplications.  One obvious
way to do this is to decompose the matrix into blocks, as in
Strassen's original algorithm.
Strassen's ``laser method''~\cite[Section~15.8]{BCS} is a
sophisticated version of this, where several matrix-multiplication
tensors are efficiently packed into a single bilinear operation (not
necessarily itself a matrix multiplication).
The rank of the tensor---in fact the border rank, which will be defined below---is used to keep
track of the complexity of the resulting recursive algorithm, and
appears in the resulting inequality for~$\omega$.
This idea of recursion is also behind the ``group-theoretic''
algorithms described in the next section.

We have mentioned that the exponent of matrix multiplication may be
defined in terms of the rank~$R\bigl(\braket{n,n,n}\bigr)$:
\[
  \omega(\F) = \inf \Set{ \tau\in\RR | R\bigl(\braket{n,n,n}\bigr)=O(n^\tau)}.
\]
The reason for dealing with the rank rather than directly with the complexity
measure is that the rank is better behaved with respect to certain
operations, and this will be useful for deriving bounds on the
asymptotic complexity via recursion.  In
particular~\cite[Proposition~14.23]{BCS}, we have
\[
  R(\vphi_1\tensor\vphi_2)\le R(\vphi_1)\tensor R(\vphi_2)
\]
for bilinear maps $\vphi_1$ and~$\vphi_2$, while the corresponding
inequality with~$L$ in place of~$R$ is not known to be true.
Let $\braket{e,h,l}$ be the tensor of
$\matrixring_{e\times h}\times \matrixring_{h\times l} \to \matrixring_{e\times l}$
matrix multiplication.
Since $\braket{e,h,l} \tensor \braket{e',h',l'} \isom
\braket{ee',hh',ll'}$~\cite[Proposition~14.26]{BCS}, we have
$R\bigl(\braket{ee',hh',ll'}\bigr)=R\bigl(\braket{e,h,l}\bigr)R\bigl(\braket{e',h',l'}\bigr)$.
Using properties of the rank function, it is easy to derive bounds
on~$\omega$ given estimates of the rank of a particular tensor.
\begin{example}
If $R\bigl(\braket{h,h,h}\bigr)\le r$, then $h^\omega\le r$.
\end{example}
\noindent
The first generalization is to allow rectangular matrices, via
symmetrization:
we have
\[R\bigl(\braket{e,h,l}\bigr) =
R\bigl(\braket{h,l,e}\bigr) = R\bigl(\braket{l,e,h}\bigr)
\]
(another nice property of the rank not shared by the
multiplicative complexity),
so if
$R\bigl(\braket{e,h,l}\bigr) \le r$, then
$R\bigl(\braket{ehl,ehl,ehl}\bigr)\le r^3$, and therefore
\begin{equation}
\label{eq:easy_inequality}
(ehl)^{\omega/3}\le r.
\end{equation}

The next refinement is to multiply several matrices at once.  But
first we need to discuss border rank.
The border rank appears as follows.
The idea is that one may be able to
approximate a tensor~$t$ of a certain rank by 
a family~$t_1(\veps)=\sum_{i=1}^ru_i(\veps)\tensor v_i(\veps)\tensor w_i(\veps)$ of tensors of possibly smaller rank,
meaning
\[
  \veps^{1-q}t_1(\veps) = t + O(\veps)
\]
for some positive integer~$q$.
The border rank~$\borderrank(t)$ is the smallest~$r$ for which this is possible.
This has a geometric interpretation, studied
by Landsberg~\cite{L}.

The border rank is
always less than or equal to the rank, and shares some of its
properties, including that of being hard to determine.
Landsberg~\cite{L2} proved that $\borderrank\bigl(\braket{2,2,2}\bigr)=
R\bigl(\braket{2,2,2}\bigr)=7$, but for~$n=3$ the best result
known is $14 \le \borderrank\bigl(\braket{3,3,3}\bigr) \le 21$
(to be compared with the estimate~$19\le R\bigl(\braket{3,3,3}\bigr)\le23$
mentioned before).

The border rank may be strictly less than the rank.
For instance, the rank of
\[
  t = x_1\tensor y_1\tensor(z_1+z_2) + x_1\tensor y_2\tensor z_1 +
  x_2\tensor y_1\tensor z_1
\]
is $3$, but its border rank is only~$2$:
\[
  \veps^{-1}t_1(\veps) \defeq \veps^{-1}\left[(\veps-1)x_1\tensor y_1\tensor z_1 + (x_1+\veps
    x_2)\tensor(y_1+\veps y_2)\tensor(z_1+\veps z_2)\right] = t + O(\veps),
\]
as can be seen by expanding the left-hand side.

The importance of the border rank is that, as in this example, the original tensor may be recovered from~$t_1(\veps)$
by computing the coefficient of some power of~$\veps$; in other
words, from such an approximate algorithm for computing~$t$ we may recover an exact one.
This expansion
increases the number of monomials, so this does not help to compute~$t$ itself;
the magic happens when we compute~$t^{\tensor N}$ for large~$N$.
Taking tensor powers corresponds to multiplying matrices recursively.

The border rank replaces the rank in a refinement of~\eqref{eq:easy_inequality},
so that $\borderrank\bigl(\braket{e,h,l}\bigr)\le r$ implies
$(ehl)^{\omega/3}\le r$.
A bit of work, generalizing this to the case of several simultaneous
matrix multiplications, results in Sch\"onhage's
\emph{asymptotic sum inequality}
\begin{equation}
\label{eq:asymptotic_sum_inequality}
  \borderrank\left(\bigoplus_{i=1}^s\braket{e_i,h_i,l_i}\right)\le r
  \implies
  \sum_{i=1}^s(e_ih_il_i)^{\omega/3}\le r.
\end{equation}

From these sorts of considerations, one can see that good bounds
on the asymptotic complexity of matrix multiplication can be obtained by
constructing specific tensors of small border rank which contain matrix
tensors as components;
this is the idea behind Strassen \textit{et al.}'s
laser method.

The principle of the laser method~\cite[Proposition~15.41]{BCS}
is to look for a tensor~$t$, of small border rank,
which has a direct-sum decomposition into blocks each
of which is isomorphic to a matrix tensor, and whose
support is ``tight'', ensuring that in a large power of~$t$
one can find a sufficiently large direct sum of matrix
tensors.  Then one can apply~\eqref{eq:asymptotic_sum_inequality}.

This combinatorial method was used by Coppersmith and Winograd~\cite{CW}
to derive $\omega < 2.376$, the best estimate currently
known.

\subsection{Group-theoretic methods of fast matrix multiplication}
\label{sec:group_theoretic_algorithms}

As explained in the previous section, the general principle is to
embed several simultaneous matrix multiplications in a single tensor,
via some combinatorial construction to ensure that the embedding is efficient.

A rough sketch of Cohn \textit{et al.}'s~\cite{CKSU} ``group-theoretic'' algorithms is that
they involve embedding matrix multiplication into multiplication in a
group algebra~$\CC[G]$ of a finite group~$G$.
The embedding uses three subsets of~$G$ satisfying the ``triple product
property'' to encode matrices as elements of the group algebra, so that
the matrix product can be read off the corresponding product in~$\CC[G]$.
The number of operations required to multiply two
matrices is, therefore, less than or equal to the number of operations
required to multiply two elements of~$\CC[G]$.
As a ring, $\CC[G] \isom \matrixring_{d_1\times
  d_1}(\CC)\times\cdots\times\matrixring_{d_r\times d_r}(\CC)$,
where $d_1,\ldots,d_r$ are the dimensions of the irreducible
representations of~$G$ (see, for instance,~\cite[Chapter~3]{Lam}).
This isomorphism may be realized as a Fourier transform on~$G$,
which can be computed efficiently.
In other words, multiplication in~$\CC[G]$ is
equivalent to several smaller matrix multiplications, and one can
apply the algorithm recursively in order to get a bound on~$\omega$.

\Cohnetal's embedding is of a very particular type, based on the following
triple product property: if there are subsets $X,\,Y,\,Z\subseteq
G$
such that $x{x'}^{-1}y{y'}^{-1}z{z'}^{-1}=1$, then $x=x'$, $y=y'$, and~$z=z'$.
This realizes the $|X|\times |Y|$ by $|Y|\times|Z|$ matrix
multiplication~$AB$ by sending $a_{xy}$ to $\sum a_{xy}x^{-1}y$
and $b_{y'z}$ to $\sum b_{y'z}y'^{-1}z$; the triple product property
ensures that one can extract the matrix product from the product in
the group algebra by looking at the coefficients of~$x^{-1}z$
for $x\in X$ and $z\in Z$.

It may be more convenient, as in the previous section, to encode
\emph{several} matrix
multiplications via the \emph{simultaneous} triple product property:
for $X_i,\,Y_i,\,Z_i\subseteq H$ one should have
$x_i{x'_j}^{-1}y_j{y'_k}^{-1}z_k{z'_i}^{-1}=1 \longrightarrow i=j=k$ and
$x_i=x'_i$, $y_i=y_i'$, $z_i=z'_i$.
It follows from~\eqref{eq:asymptotic_sum_inequality}
that
\[
\sum_i\bigl(|X_i||Y_i||Z_i|\bigr)^{\omega/3}\le\sum_{k}d_k^\omega.
\]
\noindent
We remark that the simultaneous triple product property
in~$H$ reduces to the triple product property in the
wreath product~$G=H^n\rtimes\Sym_n$, so the groups
actually output by this method turn out rather large.

From this initial description it is not at all clear what kinds of
groups will give good bounds.
To this end, Cohn~\textit{et al.}\ introduce several combinatorial
constructions, analogous to those of Coppersmith and Winograd,
which produce subsets satisfying the simultaneous triple product property
inside powers~$H^k$ of a finite Abelian group~$H$,
and hence the triple product property inside wreath products
of~$H$ with the symmetric group.
This reproduces the known bounds~$\omega<2.376$, etc.

The group-theoretic method therefore provides another perspective on
efficiently packing several independent matrix multiplications into one.
In both cases the essential problem seems to be a combinatorial one,
and one can state combinatorial conjectures which would imply~$\omega=2$.

\subsection{Asymptotic complexity of other linear problems}
\label{sec:other_linear_problems}

One can also use recursive ``divide-and-conquer'' algorithms to prove that the
asymptotic complexity of other problems in linear algebra is the same as that
of matrix multiplication.  This justifies the emphasis placed on matrix
multiplication in numerical linear algebra.

As a simple example, we will begin with
\begin{example}[matrix inversion]
On one hand, we have the identity
\[
\begin{pmatrix}
I & A & 0 \\
0 & I & B \\
0 & 0 & I
\end{pmatrix}^{-1}
=
\left( \begin{array}{rrr}
I & -A & AB \\
0 &  I & -B \\
0 &  0 &  I
\end{array} \right),
\]
which shows that two $n\times n$~matrices may be multiplied
by inverting a $3n\times 3n$~matrix.  This shows
that if an invertible $n\times n$~matrix can be inverted
in $O(n^{\omega+\eta})$~operations, then the product of two
arbitrary $n\times n$~matrices can also be computed
in $O(n^{\omega+\eta})$~operations.

In the other direction, consider the identity
\[
\begin{pmatrix}
A & B \\
C & D
\end{pmatrix}^{-1}
=
\begin{pmatrix}
A^{-1} + A^{-1}BS^{-1}CA^{-1} & -A^{-1}BS^{-1} \\
-S^{-1}CA^{-1} & S^{-1}
\end{pmatrix},
\qquad S \defeq D-CA^{-1}B.
\]
This shows that inversion of~$\left(\begin{smallmatrix}A & B\\C & D\end{smallmatrix}\right)\in M_{2n\times2n}(\CC)$ can be reduced to a certain (fixed) number of 
$n\times n$~matrix multiplications and inversions.\footnote{For instance, $2$~inversions and $6$~multiplications.}
Unfortunately, the indicated inverses, e.g., $A^{-1}$, may not
exist.  This defect may be remedied by writing
$\left(\begin{smallmatrix}A & B\\C & D\end{smallmatrix}\right)
= X = X^*(XX^*)^{-1}$.  Now $XX^*$ is a positive-definite Hermitian
matrix, to which the indicated algorithm may be applied (both its
upper-left block and its Schur complement will be positive-definite
and Hermitian).  We conclude that fast multiplication implies
fast inversion of positive-definite Hermitian,
and therefore of arbitrary (invertible), matrices.
\end{example}

\begin{example}[$\LU$~decomposition]
Suppose, for instance, that one wishes to decompose a matrix~$A$ as
$A=LUP$, where $L$ is lower triangular and unipotent, $U$ is upper
triangular, and $P$ is a permutation matrix.
Note that not every matrix has such a decomposition; a sufficient
condition for it to exist is that $A$ have full row rank.

One can give a recursive algorithm~\cite[Theorem~16.4]{BCS}, due to
Bunch and Hopcroft,
for computing the decomposition in case $A$ has full row rank, via a
$2\times2$~block decomposition of~$A$.  This involves one inversion of
a triangular matrix,
two applications of the algorithm to smaller matrices, and several
matrix multiplications; we elide the details.  Since multiplication
and inversion can be done fast, analysis of this algorithm shows that
if an $n\times n$~matrix can be multiplied in
$O(n^{\omega+\eta})$~operations, then the $\LU$~decomposition of an
$m\times n$~matrix can be done in $O(nm^{\omega+\eta-1})$~operations,
that is $O(n^{\omega+\eta})$ in the case of a square matrix.

To show, conversely, that fast $\LU$~decomposition implies fast matrix
multiplication, one notes that $\det A$ may be computed from an
$\LU$~decomposition of~$A$, and that computing determinants is at
least as hard as matrix multiplication (cf.~\cite[Theorem~16.7]{BCS}).
This shows that the exponents of matrix multiplication,
$\LU$~decomposition, and determinants coincide.\footnote{Compare this
  result on determinants with the problem of computing the permanent,
  which is $\NP$-hard!}
\end{example}

Further examples involving other linear problems may be found in the
literature; see \cite{BCS} and also
Section~\ref{sec:stable_other_linear_problems}.

\section{Numerical stability of linear problems}

Numerical stability is just as important  for the implementation of any 
algorithm as computational cost, since accumulation and propagation of 
roundoff  errors may significantly distort the output of the algorithm, 
making the algorithm essentially useless.  On the other hand, if roundoff
error bounds can be established for a given algorithm, this guarantees that 
its output values can be trusted to lie within the regions provided by the 
error bounds. Moreover, such regions can typically be made small by increasing
the hardware precision appropriately.  Fast matrix multiplication algorithms, 
from Strassen's  algorithm to the recent group-theoretic algorithms of 
\Cohnetal \/, can be  analysed in a  uniform fashion from the stability 
point of view \cite{DDHK07}.  

The roundoff-error analysis of Strassen's method was first performed by Brent
(\cite{brent3,higham90b}, see also \cite[chap. 23]{higham96}). The analysis
of subsequent Strassen-like algorithms is due a number of authors, most notably
by Bini and  Lotti~\cite{BiniLotti}. 
This latter approach was advanced in \cite{DDHK07} to build an inclusive 
framework that accommodates all Strassen-like  algorithms based on stationary 
partitioning, bilinear  algorithms with non-stationary partitioning,  
and finally the group-theoretic algorithms of the kind developed in~\cite{CU} 
and~\cite{CKSU}. 
Moreover, combining this framework with a result of Raz~\cite{Raz}, 
one can prove that there exist numerically stable matrix multiplication 
algorithms which perform $O(n^{\omega+\eta})$~operations, for arbitrarily
small~$\eta>0$, where $\omega$ is the exponent of matrix multiplication.


The starting point of the error analysis \cite{DDHK07} is the so-called \emph{classical model of 
rounded arithmetic,} where each arithmetic operation 
introduces a small multiplicative error, i.e., the computed value of each arithmetic 
operation~$\op(a,b)$ is given by $\op(a,b)(1+\theta)$ where $|\theta|$ is bounded by
some fixed {\em machine precision\/}~$\varepsilon$ but is otherwise arbitrary. The 
arithmetic operations in classical arithmetic are $\{+,-, \cdot \}$. 
The roundoff errors are assumed to be introduced by \emph{every execution} of any 
arithmetic operation. It is further assumed that all algorithms output the exact value 
in the absence of roundoff errors (i.e., when all errors~$\theta$ are zero). 

The error analysis can be performed with respect to various norms on the matrices $A$, $B$, $C=AB$,
as will be made clear in the next section. It leads to error bounds of the form 
\begin{equation} 
\| C_{\comp}-C\|\leq \mu(n) \varepsilon \|A\| \, \|B \| +O(\varepsilon^2), \label{gen_bound} \end{equation}
with $\mu(n)$ typically low-degree polynomials in the order~$n$ of the 
matrices involved, so that $\mu(n)=O(n^c)$ for some constant~$c$. 
Switching from one norm to another is always possible, using the equivalence of norms on a 
finite-dimensional space, but this may incur additional factors that depend on~$n$.

\subsection{Recursive matrix multiplication: Strassen and beyond}
\label{sec:recursive_matrix_multiplication}

In his breakthrough paper \cite{S}, Strassen observed that the multiplication
of two $2\times 2$~block matrices requires only~$7$ (instead of~$8$) block 
multiplications, and used that remarkable observation recursively to obtain
a matrix-multiplication algorithm with running time~$O(n^{\log_2 7})$.
Precisely, the product
\[\left( \begin{array}{cc} A_{11} & A_{12}  \\ A_{21} & A_{22} \end{array} \right)
\times \left( \begin{array}{cc} B_{11} & B_{12}  \\ B_{21} & B_{22}  \end{array} \right)=
\left(\begin{array}{cc} C_{11} & C_{12} \\ C_{21} & C_{22}   \end{array}  \right),
\]
can be computed by calculating the submatrices 
\begin{eqnarray*}
 M_1 & = & (A_{11}+A_{22})(B_{11}+B_{22})  \\
 M_2 & = & (A_{21}+A_{22}) B_{11} \\
 M_3 & = & A_{11} (B_{12}-B_{22}) \\
 M_4 & = & A_{22} (B_{21}-B_{11}) \\
 M_5 & = & (A_{11}+A_{12}) B_{22} \\
 M_6 & = & (A_{21}-A_{11}) (B_{11}+B_{12}) \\
 M_7 & = & (A_{12}-A_{22}) (B_{21}+B_{22})
 \end{eqnarray*}
and then combining them linearly as
\begin{eqnarray*}
 C_{11} & = & M_1+M_4-M_5+M_7  \\
C_{12} & = & M_3+M_5 \\
C_{21} & = & M_2+M_4 \\
C_{22} & = & M_1-M_2+M_3+M_6 .
 \end{eqnarray*}

Starting with matrices of dyadic order, this algorithm can be applied
by recursively partitioning each matrix into four square blocks and 
running these computations.  This yields running time 
$O(n^{\log_2 7})\approx O(n^{2.81})$.  Since any matrix can be padded
with zeros to achieve the nearest dyadic order, the dyadic size assumption
is not restrictive at all.

The breakthrough of Strassen generated a flurry of activity in the area,
leading to a number of subsequent improvements, among those by Pan~\cite{Pan},
Bini \emph{et al.}~\cite{Bini},  
Sch\"onhage~\cite{Schoen}, Strassen~\cite{S2}, and eventually Coppersmith and 
Winograd~\cite{CW}.  Each of these algorithms is Strassen-like, i.e., uses recursive 
partitioning and a special ``trick'' to reduce the number of block matrix 
multiplications.

Such recursive algorithms for matrix multiplication can be analysed as follows. 
Recall that bilinear functions can be evaluated via bilinear algorithms,
as in Equation~\eqref{eq:bilinear_rank}.
Since they do not use commutativity of the coordinates, these algorithms
apply equally well when the input entries are elements of a non-commutative
algebra; their recursive use for matrix multiplication is then straightforward.
A \emph{bilinear non-commutative algorithm} (see \cite{BiniLotti} or~\cite{BrockettDobkin}) 
that computes products~$C=AB$ in~$\matrixring_{k\times k}(\F)$
using $t$~non-scalar multiplications over a subfield~$\H \subseteq \F$
(not necessarily equal to~$\F$)\footnote{Field extensions have no effect on the asymptotic complexity, but changing~$\H$ will affect the constants in~$\mu(n)$.} is determined by 
three  $k^2{\times}t$ matrices $U$, $V$ and~$W$ with elements in~$\H$ 
such that
\begin{equation} 
c_{hl}=\sum_{s=1}^t w_{rs} P_s, \;\; {\rm where} \;\; 
P_s =  \left(\sum_{i=1}^{k^2} u_{is} x_i \right) \left(\sum_{j=1}^{k^2} v_{js} y_j \right),
 \;\;\;\;  \begin{array}{l} r=k(h-1)+l,\\ h,l=1,\ldots,k, \end{array}   \label{main} 
\end{equation}
where $x_i$ (resp. $y_j$) are the elements of $A=(a_{ij})$ (resp. of $B=(b_{ij})$) ordered column-wise, 
and $C=(c_{ij})$ is the product~$C=AB$.  

For an arbitrary~$n$, the algorithm consists of recursive partitioning and 
applying~(\ref{main}) to compute products of resulting block matrices. More precisely,
suppose that $A$ and~$B$ are of size~$n{\times}n$, where $n$ is a power of~$k$ (which
can always be achieved by padding the matrices $A$ and~$B$ with zero columns and rows,
as we already mentioned). Partition $A$ and~$B$ into $k^2$~square blocks $A_{ij}$,~$B_{ij}$ of size~$(n/k){\times}(n/k)$.
Then the blocks~$C_{hl}$ of the product~$C=AB$ can be computed by applying~(\ref{main}) to 
the blocks of $A$ and~$B$, where each block $A_{ij}$,~$B_{ij}$ has to be again partitioned 
into $k^2$~square sub-blocks to compute the $t$ products~$P_s$ and then the blocks~$C_{hl}$.
The algorithm obtained by running this recursive procedure $\log_k n$~times computes the
product~$C=AB$ using at most $O(n^{\log_k t})$~multiplications.

\begin{theorem}[\protect{\cite[Theorem~3.1]{DDHK07}}] \label{thm:stationary}
A bilinear non-commutative algorithm for matrix multiplication based
on stationary partitioning is stable. It satisfies the error bound~(\ref{gen_bound}) 
where $\|\cdot\|$ is the maximum-entry norm and where
\[
 \mu(n) = \bigl(1+\max_{r,s}(\alpha_s+\beta_s+\gamma_r+3)
\log_k n\bigr) \cdot \bigl( \om \cdot \|U\|  \,\|V\| \, \|W\| \bigr)^{\log_k n}. 
\]
Here $\alpha_s=\lceil\log_2 a_s\rceil$, $\beta_s=\lceil\log_2 b_s\rceil$ and~$\gamma_r = \lceil \log_2 c_r\rceil$
where $a_s$ and $b_s$ (resp.~$c_r$) are the number of non-zero entries of $U$ and $V$ (resp.~$W$) in column~$s$ (resp. row~$r$),
while
$\om$ is an integer that depends (in a rather involved way) on the sparsity pattern
of the matrices $U$, $V$ and~$W$.
\end{theorem}

\noindent
This theorem can be subsequently combined with the result of Raz \cite{Raz} 
that the exponent of matrix multiplication is achieved by bilinear
non-commutative algorithms~\cite{Raz} to produce an important corollary:

\begin{corollary}[\protect{\cite[Theorem~3.3]{DDHK07}}] \label{thm:omega-is-stable}
For every $\eta>0$ there exists an algorithm
for multiplying $n$-by-$n$ matrices which performs
$O(n^{\omega+\eta})$ operations (where
$\omega$ is the exponent of matrix multiplication)
and which is numerically stable, in the sense that it 
satisfies the error bound~(\ref{gen_bound}) with 
$\mu(n)=O(n^c)$ for some constant $c$ depending on
$\eta$ but not $n$.
\end{corollary}

\noindent
The analysis of stationary algorithms extends to bilinear matrix multiplication 
algorithms based on non-stationary partitioning. This means that the matrices $A^{[j]}_{s,\comp}$ 
and~$B^{[j]}_{s,\comp}$ are partitioned into $k{\times}k$ square blocks, but $k$ depends on
the level  of recursion, i.e., $k=k(j)$, and the corresponding matrices $U$, $V$ and~$W$ also
depend on $j$: $U=U(j)$, $V=V(j)$,~$W=W(j)$.   Otherwise the algorithm proceeds exactly like
the stationary algorithms.

Finally, algorithms that combine recursive non-stationary partitioning with pre- and
post-processing given by linear maps $\pre_n()$ and~$\post_n()$ acting on matrices of an 
arbitrary order~$n$ can be analysed using essentially the same approach~\cite{DDHK07}.
Suppose that the matrices $A$ and~$B$ are each (linearly) pre-processed, then partitioned into 
blocks, respective pairs of blocks are multiplied recursively and assembled into a 
large matrix, which is then (linearly) post-processed to obtain the resulting matrix~$C$.

The analysis in \cite{DDHK07} is performed for an arbitrary  \emph{consistent} (i.e., 
submultiplicative) norm~$\| \cdot \|$ that in addition must be  defined for  matrices of 
all sizes and must satisfy the condition
\begin{equation}
\max_s  \|M_{s} \|\leq  \|M \|\leq \sum_{s}\|M_{s} \|
\label{normcond}
\end{equation}
 whenever the matrix $M$ is partitioned into
 blocks $(M_{s})_s$ (an example of such a norm is provided by the $2$-norm $\|\cdot \|_2$). 
Note that the previously mentioned maximum-entry norm satisfies~(\ref{normcond}) but is not 
consistent, i.e., does not satisfy
$$ \| A B\| \leq \|A \| \cdot \|B \| \qquad \hbox{\rm for all } A, \; B.$$ 

Denoting the norms of pre- and post- processing maps subordinate to
the norm $\| \cdot \|$ by $\| \cdot \|_{\op}$, we suppose  that the pre-
 and post-processing is performed with errors 
\begin{eqnarray*}
 \| \pre_n (M)_{\comp}-\pre_n (M) \|_{\op} & \leq & f_{\text{pre}}(n) \ve \|M\|+ O(\varepsilon^2), \\
\| \post_n (M)_{\comp}-\post_n (M) \|_{\op} & \leq & f_{\text{post}}(n) \ve \|M\|+ O(\varepsilon^2), 
\end{eqnarray*}
where $n$ is the order of the matrix~$M$. 
As before, we denote by $\mu(n)$ the coefficient of~$\varepsilon$ in the final error 
bound~(\ref{gen_bound}). 

Under all these assumptions, the following error estimate follows:
\begin{theorem}[\protect{\cite[Theorem~3.5]{DDHK07}}] \label{thm_crude}
A recursive matrix multiplication algorithm based on 
non-stationary  partitioning with pre- and post-processing is stable. It satisfies the error 
bound~(\ref{gen_bound}), with the function~$\mu$ satisfying the recursion
$$  \mu(n_j) = \mu(n_{j+1}) t_j \|\post_{n_j} \|_{\op} \, \|\pre_{n_j} \|_{\op}^2 + 2  f_{\text{pre}}(n_j) t_j \|\post_{n_j}\|_{\op} 
+f_{\text{post}}(n_j) \|\pre_{n_j} \|_{\op}^2$$  for $ j=1, \ldots, p.  $
\end{theorem}

\subsection{Group-theoretic matrix multiplication}

In this section we describe the group-theoretic constructions of \Cohnetal \/
Our exposition closely follows the pertinent parts of \cite{DDHK07}. 
To give a general idea about group-theoretic fast matrix multiplication,
we must first recall some basic definitions from algebra.
\begin{definition}[semidirect product]
If $H$ is any group and $Q$ is a group which
acts (on the left) by automorphisms of $H$, with 
$q \cdot h$ denoting
the action of $q \in Q$ on $h \in H$, then 
the \emph{semidirect product} $H \rtimes Q$ is the
set of ordered pairs $(h,q)$ with the
multiplication law
\begin{equation} \label{eqn:semidirect}
(h_1,q_1) (h_2,q_2) = (h_1(q_1 \cdot h_2), q_1 q_2).
\end{equation}
We will identify $H \times \{1_Q\}$ with $H$
and $\{1_H\} \times Q$ with $Q$, so that 
an element $(h,q) \in H \rtimes Q$ may
also be denoted simply by $hq$.  Note that
the multiplication law of $H \rtimes Q$
implies the relation $qh = (q \cdot h) q$.
\end{definition}

\begin{definition}[wreath product]
If $H$ is any group, $S$ is any finite set,
and $Q$ is a group with a left action on $S$,
the \emph{wreath product} $H \wr Q$ is the semidirect
product $(H^S) \rtimes Q$ where $Q$ acts on 
the direct product of $|S|$~copies of~$H$
by permuting the coordinates according to 
the action of $Q$ on~$S$.  (To be more
precise about the action of $Q$ on~$H^S$,
if an element $h \in H^S$ is represented
as a function~$h\colon S \rightarrow H$, then 
$q \cdot h$ represents the function 
$s \mapsto h(q^{-1}(s)).$)
\end{definition}

\begin{definition}[triple product property,
simultaneous triple product property]
If $H$ is a group and $X,\,Y,\,Z$ are three subsets,
we say $X,\,Y,\,Z$ satisfy the \emph{triple product
property} if it is the case that for all
$q_x \in Q(X)$, $q_y \in Q(Y)$, $q_z \in Q(Z)$, if
$ q_x q_y q_z = 1 $
then $q_x=q_y=q_z=1$.
Here $Q(X)=Q(X,X)$ is the set of quotients;
$Q(S,T)\defeq\Set{ st^{-1} | s\in S,\;t\in T}\subseteq H$.

If $\Set{(X_i,Y_i,Z_i) | i \in I }$ is a 
collection of ordered triples of subsets of~$H$,
we say that this collection satisfies
the \emph{simultaneous triple product property (STPP)}
if it is the case that for all $i,\,j,\,k \in I$ and all
$q_x \in Q(X_i,X_j)$, $q_y \in Q(Y_j,Y_k)$, $q_z \in Q(Z_k,Z_i)$, if
$q_x q_y q_z = 1$ then
$q_x = q_y = q_z = 1$ and $i=j=k$.
\end{definition}

\begin{definition}[Abelian STP family]
\label{def_stpp_alg}
An \emph{Abelian STP family} with growth parameters~$(\alpha,\beta)$
is a collection
of ordered triples~$(H_N,\Upsilon_N, k_N)$, defined
for all~$N > 0$, satisfying
\begin{enumerate}
\item $H_N$ is an Abelian group.
\item 
$\Upsilon_N = \Set{(X_i,Y_i,Z_i) | i = 1,2,\ldots,N}$
is a collection of $N$ ordered triples of subsets
of~$H_N$ satisfying the simultaneous triple
product property.
\item
$|H_N| = N^{\alpha + o(1)}$.
\item
$k_N = \prod_{i=1}^N |X_i| =
\prod_{i=1}^N |Y_i| =
\prod_{i=1}^N |Z_i| =
N^{\beta N + o(N)}$.
\end{enumerate}
\end{definition}

Recall from Section~\ref{sec:group_theoretic_algorithms}
that in~\cite{CKSU} matrix-multiplication algorithms are
constructed based on families of wreath products of Abelian
groups.

To get into more details, we must recall basic facts about
the discrete Fourier transform of an Abelian group.  For an 
Abelian group~$H$, let $\widehat{H}$ denote the set of all 
homomorphisms from $H$ to~$S^1$, the multiplicative group of 
complex numbers with unit modulus. Elements of~$\widehat{H}$ 
are called \emph{characters} and are usually denoted by the 
letter~$\chi$. The sets $H,\,\widehat{H}$ have the same cardinality.  
When $H_1,\, H_2$ are two Abelian groups, there is a canonical 
bijection between the sets $\widehat{H_1} \times \widehat{H_2}$
and $(H_1 \times H_2)^{\wedge}$; this bijection
maps an ordered pair $(\chi_1,\chi_2)$ to the
character $\chi$ given by the formula 
$\chi(h_1,h_2) = \chi_1(h_1) \chi_2(h_2).$
Just as the symmetric group $\Sym_n$ acts on $H^n$
via the formula $\sigma \cdot (h_1,h_2,\ldots,h_n)
= (h_{\sigma^{-1}(1)},h_{\sigma^{-1}(2)},\ldots,
h_{\sigma^{-1}(n)}),$ there is a left action of
$\Sym_n$ on the set $\widehat{H}^n$ defined by the
formula $\sigma \cdot (\chi_1,\chi_2,\ldots,\chi_n)
= (\chi_{\sigma^{-1}(1)},\chi_{\sigma^{-1}(2)},\ldots,
\chi_{\sigma^{-1}(n)}).$  

\begin{notation}
The notation $\Xi(H^n)$ will be used to denote a
subset of~$\widehat{H}^n$ containing exactly one
representative of each orbit of the $\Sym_n$ action
on~$\widehat{H}^n$.  An orbit of this action
is uniquely determined by a multiset consisting of 
$n$~characters of~$H$, so the cardinality of 
$\Xi(H^n)$ is equal to the number of such multisets,
i.e. $\binom{|H|+N-1}{N}.$
\end{notation}

Given an Abelian STP family, the corresponding recursive matrix 
multiplication algorithm is defined  as follows.  Given 
a pair of $n$-by-$n$ matrices $A,B$, find the minimum $N$ 
such that  $k_N \cdot N! \ge n$, and denote the group
$H_N$ by $H$.  If $N! \ge n$, multiply the matrices using an arbitrary
algorithm.  (This is the base of the recursion.)
Otherwise reduce the problem of computing the matrix product $AB$
to $\binom{|H|+N-1}{N}$ instances of  $N! \times N!$ matrix multiplication,
using a reduction based on the discrete Fourier transform of the Abelian 
group $H^N$.

Padding the matrices with additional rows and columns of $0$'s if necessary, 
one may assume that $k_N \cdot N! = n$. Define subsets $X,Y,Z \subseteq H \wr \Sym_N$
as
$$
X  =  \left( \prod_{i=1}^N X_i \right) \times \Sym_N, \qquad
Y  =  \left( \prod_{i=1}^N Y_i \right) \times \Sym_N, \qquad
Z  =  \left( \prod_{i=1}^N Z_i \right) \times \Sym_N. 
$$
These subsets satisfy the triple product  property~\cite{CKSU}.
Note that $|X|=|Y|=|Z|=n.$  Now treat the rows
and columns of $A$ as being indexed by the sets 
$X,\,Y$, respectively; treat the rows and 
columns of~$B$ as being indexed by the sets $Y,\,Z$,
respectively.

The algorithm uses two auxiliary vector
spaces $\C[H \wr \Sym_N]$ and $\C[\widehat{H}^N \rtimes \Sym_N]$, 
each of dimensionality $|H|^N N!$ and each with a specific
basis:  the basis for  $\C[H \wr \Sym_N]$ is denoted 
by $\set{\mathbf{e}_g | g \in H \wr \Sym_N}$,
and the basis for $\C[\widehat{H}^N \rtimes \Sym_N]$ is denoted
by $\set{\mathbf{e}_{\chi,\sigma} | \chi \in \widehat{H}^N,\;
\sigma \in \Sym_N}$.

The Abelian STP algorithm from \cite{CKSU} performs the following
steps, which will be labelled according to whether they 
perform arithmetic or not. (For example, a permutation of
the components of a vector does not involve any arithmetic.)
\begin{enumerate}
\item   \label{step:embed}
\textbf{Embedding}
\textsc{(no arithmetic):\hspace{5mm}}
Compute the following pair of vectors in
$\C[H \wr \Sym_N]$.
\begin{eqnarray*}
a & \eqbd & \sum_{x \in X} \sum_{y \in Y} 
A_{xy} \mathbf{e}_{x^{-1} y} \\
b & \eqbd & \sum_{y \in Y} \sum_{z \in Z} 
B_{yz} \mathbf{e}_{y^{-1} z}. 
\end{eqnarray*}
\item   \label{step:fourier}
\textbf{Fourier transform}
\textsc{(arithmetic):\hspace{5mm}}
Compute the following pair of vectors in
$\C[\widehat{H}^N \rtimes \Sym_N]$.
\begin{eqnarray*}
\hat{a} & \eqbd &
\sum_{\chi \in \widehat{H}^N}
\sum_{\sigma \in \Sym_N} \left(
\sum_{h \in H^N} \chi(h) a_{\sigma h} 
\right) \mathbf{e}_{\chi, \sigma}. \\
\hat{b} & \eqbd &
\sum_{\chi \in \widehat{H}^N}
\sum_{\sigma \in \Sym_N} \left(
\sum_{h \in H^N} \chi(h) b_{\sigma h} 
\right) \mathbf{e}_{\chi, \sigma}. 
\end{eqnarray*}
\item   \label{step:assemble}
\textbf{Assemble matrices}
\textsc{(no arithmetic):\hspace{5mm}}
For every $\chi \in \Xi(H^N)$,
compute the following pair of matrices
$A^\chi, 
B^\chi 
$, whose rows and columns
are indexed by elements of $\Sym_N$.
\begin{eqnarray*}
A^\chi_{\rho \sigma} & \eqbd & \hat{a}_{\rho \cdot \chi, 
\sigma \rho^{-1}} \\
B^\chi_{\sigma \tau} & \eqbd & \hat{b}_{\sigma \cdot \chi,
\tau \sigma^{-1}} \\
\end{eqnarray*}
\item  \label{step:multiply}
\textbf{Multiply matrices}
\textsc{(arithmetic):\hspace{5mm}}
For every $\chi \in \Xi(H^N)$,
compute the matrix product $C^\chi \eqbd A^\chi B^\chi$
by recursively applying the Abelian STP algorithm.
\item  \label{step:disassemble}
\textbf{Disassemble matrices}
\textsc{(no arithmetic):\hspace{5mm}}
Compute a vector $\hat{c} \eqbd
 \sum_{\chi,\sigma} \hat{c}_{\chi,\sigma} \mathbf{e}_{\chi,\sigma}
\in \C[\widehat{H}^N \rtimes \Sym_N]$
whose components $\hat{c}_{\chi,\sigma}$ are defined as follows.
Given $\chi,\sigma,$ let $\chi_0 \in \Xi(H^N)$ and $\tau \in
\Sym_N$ be such that $\chi = \tau \cdot \chi_0.$  Let
$$
\hat{c}_{\chi,\sigma} \eqbd C^{\chi_0}_{\tau,\sigma \tau}.
$$
\item  \label{step:inverseFT}
\textbf{Inverse Fourier transform}
\textsc{(arithmetic):\hspace{5mm}}
Compute the following vector $c \in \C[H \wr \Sym_N]$.
$$
c \eqbd \sum_{h \in H^N} \sum_{\sigma \in \Sym_N} \left(
\frac{1}{|H|^N} \sum_{\chi \in \widehat{H}^N} \chi(-h)
\hat{c}_{\chi,\sigma} \right) \mathbf{e}_{\sigma h}.
$$
\item  \label{step:output}
\textbf{Output}
\textsc{(no arithmetic):\hspace{5mm}}
Output the matrix $C = (C_{xz})$ whose entries are 
given by the formula $$C_{xz} \eqbd c_{x^{-1}z}.$$
\end{enumerate}

\noindent
The main result of \cite{DDHK07} establishes the numerical stability
of all Abelian STP algorithms.

\begin{theorem}[\protect{\cite[Theorem~4.13]{DDHK07}}] \label{thm_error_analysis}
If $\Set{(H_N,\Upsilon_N,k_N ) }$ is an Abelian
STP family with growth parameters $(\alpha,\beta)$,
then the corresponding Abelian STP algorithm
is stable. It satisfies the error bound~(\ref{gen_bound}), 
with the Frobenius norm and the function $\mu$ of order 
$$  \mu(n)=n^{\frac{\alpha+2}{2\beta} \,+\, o(1)}. $$
\end{theorem}

\begin{remark}[\protect{\cite[Remark~4.15]{DDHK07}}]
The running time of an Abelian STP algorithm can also be
bounded in terms of the growth parameters of the Abelian
STP family.  Specifically, the running time is  \cite{CKSU} 
$O \left( n^{(\alpha-1)/\beta \, + \, o(1)} \right).$
Note the curious interplay between the two exponents,
 $(\alpha-1)/\beta$ and $(\alpha+2)/2\beta$: their sum is 
always bigger than $3$, since $\alpha \geq 2\beta+1$ is 
one of the requirements for an Abelian STP construction: 
$$ {\alpha -1\over \beta} + {\alpha +2 \over 2\beta}=
{3\alpha\over 2\beta} \geq {6\beta+3\over 2\beta}>3 .$$ 
\end{remark}

\subsection{Matrix decompositions and other linear problems}
\label{sec:stable_other_linear_problems}

The  results about matrix multiplication from the previous section 
can be extended to show that essentially {\em all}
linear algebra operations can also be done stably, in time $O(n^{\omega})$
or~$O(n^{\omega + \eta})$, for arbitrary~$\eta > 0$~\cite{DDH07}.
For simplicity, whenever an exponent contains ``$+ \eta$'', it 
will henceforth mean ``for any $\eta > 0$''.  Below we summarize 
the main results of \cite{DDH07}.

The first result in \cite{DDH07} can be roughly summarized by saying 
that $n$-by-$n$ matrices
can be multiplied in $O(n^{\omega + \eta})$~operations 
\emph{if and only if} $n$-by-$n$ matrices can be inverted stably 
in $O(n^{\omega + \eta})$~operations.
Some extra precision is necessary to make this claim;
the cost of extra precision is included in the $O(n^{\eta})$ factor.

Other results in~\cite{DDH07}  may be summarized by saying that if $n$-by-$n$ matrices
can be multiplied in $O(n^{\omega + \eta})$ {\em arithmetic} operations, 
then the QR~decomposition can be computed stably  (moreover, linear systems 
and least squares problems can be solved stably) in $O(n^{\omega + \eta})$ 
\emph{arithmetic} operations. 
These results do not require extra precision, which is why one 
needs to count arithmetic operations rather than bit operations.

The QR~decomposition can be used to stably compute a rank-revealing
decomposition, the (generalized) Schur form,
and  the singular value decomposition, 
all in $O(n^{\omega + \eta})$ \emph{arithmetic} operations. 
Computing (generalized) eigenvectors from the Schur form,
can be done by solving the (generalized) Sylvester equation,
all of which can be done stably in $O(n^{\omega + \eta})$ 
\emph{bit} operations.

Here are a few more details about the work in~\cite{DDH07}.
The paper starts off by reviewing conventional  block algorithms used  
in libraries like LAPACK \cite{lapackmanual3} and ScaLAPACK \cite{scalapackmanual}.
The normwise backward stability of these algorithms was shown earlier 
\cite{higham90b,demmelhighamschreiber,higham96}
using~(\ref{gen_bound}) as an assumption. This means that these
algorithms are guaranteed to produce the exact answer (e.g., solution
of a linear system) for a matrix~$\hat{C}$ close to the actual 
input matrix~$C$, where close means close in norm: 
$$\|\hat{C} - C \| = O(\varepsilon) \|C\|.$$ Here the $O(\varepsilon)$ is
interpreted to include a factor~$n^c$ for a modest constant~$c$.

The running-time analysis of these block algorithms in \cite{DDH07}
shows that these block algorithms run only as fast as 
$O(n^{\frac{9 - 2\gamma}{4 - \gamma}})$~operations, 
where $O(n^{\gamma})$ is the operation count of matrix multiplication,
with $\gamma$ used instead of $\omega + \eta$ to simplify notation.
Even if $\gamma$ were to drop from 3 to~2, the exponent $\frac{9 - 2\gamma}{4 - \gamma}$
would only drop from 3 to~2.5, providing only a partial improvement. 
However, further results in \cite{DDH07} demonstrate that one can do better.

The next step in~\cite{DDH07} is the application of known divide-and-conquer 
algorithms for reducing the complexity of matrix inversion to the complexity 
of matrix multiplication. These algorithms are not backward stable in the 
conventional sense. However,  they can be shown to achieve the same
forward error bound (bound on the norm of the error in the output) as
a conventional backward stable algorithm, provided that they use
just $O(\log^p n)$~times as many bits of precision in each arithmetic
operation (for some $p>0$) as a conventional algorithm. 
Such algorithms are called  \emph{logarithmically stable.} 

Incorporating the cost of this extra precise arithmetic in the analysis 
only increases the total cost by a factor at most~$\log^{2p} n$. Therefore, 
if there are matrix multiplication
algorithms running in $O(n^{\omega + \eta})$~operations for any $\eta > 0$,
then these logarithmically stable algorithms for operations like
matrix inversion also run in $O(n^{\omega + \eta})$ operations 
for any $\eta > 0$, and satisfy the same error bound as a conventional algorithm.

A divide-and-conquer algorithm  for QR decomposition from~\cite{ElmrothGustavson2000} 
is simultaneously  backward stable in the  conventional normwise sense (i.e., without 
extra precision),  and runs in  $O(n^{\omega + \eta})$ operations for any $\eta > 0$. 
This algorithm may be in turn used to solve linear systems, least-squares problems, 
and compute determinants equally stably and fast. The same idea applies to LU~decomposition 
but stability depends on a particular pivoting assumption~\cite{DDH07}.

The QR decomposition can then be used to compute a rank-revealing
$URV$ decomposition of a matrix~$A$. This means that $U$ and~$V$ are orthogonal,
$R$ is upper triangular, and $R$ reveals the rank of~$A$ in the following sense:
Suppose
$\sigma_1 \geq \cdots \geq \sigma_n$ are the singular values of~$A$. Then for 
each~$r$, \ $\sigma_{\min}(R(1:r,\,1:r))$ is an approximation of~$\sigma_r$
and
$\sigma_{\max} (R(r+1:n,\,r+1:n))$ is an approximation of~$\sigma_{r+1}$.
The algorithm in~\cite{DDH07} is \emph{randomized,} in the sense 
that the approximations of $\sigma_r$ and~$\sigma_{r+1}$ are reasonably accurate 
with high probability.

Finally,  the QR and URV decompositions in algorithms for the (generalized) Schur 
form of nonsymmetric matrices (or pencils) \cite{baidemmelgu94} lower their 
complexity to $O(n^{\omega + \eta})$ arithmetic operations while maintaining 
normwise backward stability.
The singular-value decomposition may in turn be reduced to solving 
an eigenvalue problem with the same complexity.
Computing (generalized) eigenvectors can only be done in a logarithmically stable
way from the (generalized) Schur form.
This is done by providing a logarithmically stable algorithm for solving the
(generalized) Sylvester equation, and using this to compute eigenvectors.

This covers  nearly all standard dense linear algebra operations
(LU decomposition, QR decomposition, matrix inversion, 
linear equation solving,
solving least squares problems, computing the (generalized) Schur form,
computing the SVD, and solving (generalized) Sylvester equations) 
and shows that all those problems can be solved stably and 
asymptotically as fast 
as the fastest matrix multiplication algorithm that may ever 
exist (whether the fastest matrix multiplication
algorithm is stable or not).  
For all but matrix inversion and solving (generalized) Sylvester equations,
stability means backward stability in
a normwise sense, and the  complexity is measured by
the usual count of arithmetic operations.

\bibliography{refs,biblio}

\end{document}